\documentstyle[12pt]{article}

 \catcode`\@=11

\ifcase\@ptsize
 \font\tenmsa=msxm10
 \font\sevenmsa=msxm7
 \font\fivemsa=msxm5
 \font\tenmsb=msym10
 \font\sevenmsb=msym7
 \font\fivemsb=msym5
\or
 \font\tenmsa=msxm10 scaled \magstephalf
 \font\sevenmsa=msxm8
 \font\fivemsa=msxm6
 \font\tenmsb=msym10 scaled \magstephalf
 \font\sevenmsb=msym8
 \font\fivemsb=msym6
\or
 \font\tenmsa=msxm10 scaled \magstep1
 \font\sevenmsa=msxm8
 \font\fivemsa=msxm6
 \font\tenmsb=msym10 scaled \magstep1
 \font\sevenmsb=msym8
 \font\fivemsb=msym6
\fi

\newfam\msafam
\newfam\msbfam
\textfont\msafam=\tenmsa  \scriptfont\msafam=\sevenmsa
  \scriptscriptfont\msafam=\fivemsa
\textfont\msbfam=\tenmsb  \scriptfont\msbfam=\sevenmsb
  \scriptscriptfont\msbfam=\fivemsb

\def\hexnumber@#1{\ifnum#1<10 \number#1\else
 \ifnum#1=10 A\else\ifnum#1=11 B\else\ifnum#1=12 C\else
 \ifnum#1=13 D\else\ifnum#1=14 E\else\ifnum#1=15 F\fi\fi\fi\fi\fi\fi\fi}

\def\msa@{\hexnumber@\msafam}
\def\msb@{\hexnumber@\msbfam}

\def\Bbb{\ifmmode\let\next\Bbb@\else
 \def\next{\errmessage{Use \string\Bbb\space only in math mode}}\fi\next}
\def\Bbb@#1{{\Bbb@@{#1}}}
\def\Bbb@@#1{\fam\msbfam#1}

\if@twoside
\else
\fi
\ifcase \@ptsize
\or 
\or 
\fi

\def\@citex[#1]#2{%
\if@filesw \immediate \write \@auxout {\string \citation {#2}}\fi
\@tempcntb\m@ne \let\@h@ld\relax \def\@citea{}%
\@cite{%
  \@for \@citeb:=#2\do {%
    \@ifundefined {b@\@citeb}%
      {\@h@ld\@citea\@tempcntb\m@ne{\bf ?}%
      \@warning {Citation `\@citeb ' on page \thepage \space undefined}}%
      {\@tempcnta\@tempcntb \advance\@tempcnta\@ne%
      \@tempcntb\number\csname b@\@citeb \endcsname \relax%
      \ifnum\@tempcnta=\@tempcntb 
        \ifx\@h@ld\relax%
          \edef \@h@ld{\@citea\csname b@\@citeb\endcsname}%
        \else%
          \edef\@h@ld{\ifmmode{-}\else--\fi\csname b@\@citeb\endcsname}%
        \fi%
      \else
        \@h@ld\@citea\csname b@\@citeb \endcsname%
        \let\@h@ld\relax%
      \fi}%
    \def\@citea{,\penalty\@highpenalty\,}%
  }\@h@ld
}{#1}}
\@addtoreset{equation}{section}
\def\section{\@startsection {section}{1}{\z@}{-3.5ex plus -1ex minus
 -.2ex}{2.3ex plus .2ex}{\large\bf\centering}}
\def\subsection{\@startsection{subsection}{2}{\z@}{-3.25ex plus -1ex minus
 -.2ex}{1.5ex plus .2ex}{\sc}}
\def\topspace{\vphantom{\vrule height 3ex depth 0pt}}
\def\bottomspace{\vphantom{\vrule height 0pt depth 2ex}}
\gdef\@publabel{\hfil}
\gdef\@pubdate{\null}
\gdef\@pubnumber{\null}
\gdef\@author{\null}
\gdef\@title{\null}
\gdef\@abstract{\null}
\long\def\pubdate#1{\gdef\@pubdate{#1}}
\long\def\pubnumber#1{\gdef\@pubnumber{#1}}
\long\def\publabel#1{\gdef\@publabel{#1}}
\long\def\author#1{\gdef\@author{#1}}
\long\def\title#1{\gdef\@title{#1}}
\long\def\abstract#1{\gdef\@abstract{#1}}
\def\titlerelax{
\let\maketitle\relax
\let\settitleparameters\relax
\let\consolidatetitle\relax
\let\inittitlepage\relax
\let\finishtitlepage\relax
\let\titlepagecontents\relax
\let\multithanks\relax
\let\titlebaselines\relax
\let\@makepub\relax
\let\@maketitle\relax
\let\@makeauthor\relax
\let\@makeabstract\relax
\let\@maketitlenote\relax
\let\thanks\relax
\let\titlerelax\relax}
\def\titleclean
{\gdef\@titlenote{}
\gdef\@abstract{}
\gdef\@author{}
\gdef\@title{}
\gdef\@pubdate{}\gdef\@pubnumber{}\gdef\@publabel{}
\gdef\@dpublabel{}
}

\def\@makepub{\vbox to \z@{\hbox to \textwidth{\hfill
\@publabel \hfill
\llap{\parbox[t]{0.25\textwidth}{\raggedleft\@pubnumber}}}%
\vss}}
\def\@maketitle{\vskip 60pt \begin{center}
 {\LARGE \@title \par}
 \end{center}}
\def\@makeauthor{{%
\def\and{\smallskip {\normalsize \rm and\smallskip }}
\def\And{\medskip {\normalsize \rm and\\}\medskip}
\long\def\address##1{{\def\and{\\and\\}\medskip
				{\small \it \\##1\\}
}}
{\centering
 \vskip 3em
 \large \lineskip .75em
 \@author}
 \par}}
\def\@makedate{\vskip 1.5em
 {\raggedright \small \noindent\@pubdate \par}}
\def\@makeabstract{\vskip 1.5em
{\small
\begin{center}
{\bf ABSTRACT\vspace{-.5em}\vspace{0pt}}
\end{center}
\quotation \@abstract \endquotation}}
\def\maketitle{\titlepage
\let\footnotesize\small \setcounter{page}{0}
\@makepub
\vfil
\@maketitle
\@makeauthor
\vfil
\@makeabstract
\@thanks
\vfil
\@makedate
\if@restonecol\twocolumn \else \eject \fi
\titlerelax \titleclean
\setcounter{footnote}{0}
}

\catcode`\@=12

\begin{document}
\bibliographystyle{npb}


\let\b=\beta
\def\blank#1{}

\def\cdd{{\cdot}}
\def\cev#1{\langle #1 \vert}
\def\cH{{\cal H}}
\def\comm#1#2{\bigl [ #1 , #2 \bigl ] }
\def\compact{ reductive}
\def\cont{\nonumber\\*&&\mbox{}}
\def\cO{{\cal O}}
\def\cul #1,#2,#3,#4,#5,#6.{\left\{ \matrix{#1&#2&#3\cr #4&#5&#6} \right\}}

\def\dz{Dz}
\def\dz{\hbox{$d\kern-1.1ex{\raise 3.5pt\hbox{$-$}}\!\!z$}}
\def\dz{ \frac{d\!z}{2\pi i}}
\def\en{\end{equation}}
\def\enn{\end{eqnarray}}
\def\eq{\begin{equation}}
\def\eqq{\begin{eqnarray}}



\def\half#1{\frac {#1}{2}}

\def\ip#1#2{\langle #1,#2\rangle}


\def\k{k}


\def\Mf#1{{M{}^{{}_{#1}}}}
\def\mno{{\textstyle {\circ\atop\circ}}}
\def\mod#1{\vert #1 \vert}

\def\Nf#1{{N{}^{{}_{#1}}}}
\def\ni{\noindent}
\def\no{{\textstyle {\times\atop\times}}}
\def\no:#1:{\mno#1\mno}
\def\nox{{\scriptstyle{\times \atop \times}}}


\let\p=\phi
\def\posdef{ positive-definite}
\def\posdefness{ positive-definiteness}
\def\Qf#1{{Q{}^{{}_{#1}}}}
\def\Qstar{\mathop{\no:QQ:}\nolimits}

\def\reductive#1#2{#1}


\def\tr{\mathop{\rm tr}\nolimits}
\def\Tr{\mathop{\rm Tr}\nolimits}







\def\vec#1{\vert #1 \rangle}
\def\vac{\vec 0}

\def\wan{$\WA_n$ }
\def\Wb{\bar W}
\def\Wf#1{{W{}^{{}_{#1}}}}
\def\wbn{$\WB_n$ }
\def\WA{\mathop{\it WA}\nolimits}
\def\WB{\mathop{\it WB}\nolimits}
\def\WBC{\mathop{\it WBC}\nolimits}
\def\WD{\mathop{\it WD}\nolimits}
\def\WG{\mathop{\it WG}\nolimits}



\def\zz#1{(z-z')^{#1}}

\openup 1\jot

\pubnumber{ EFI 92-09}
\pubdate{Feb. 1992}

\title{ Exceptional Superconformal Algebras}

\author{
P. BOWCOCK\thanks{ Email \tt
Bowcock@rabi.uchicago.edu}
\thanks{Supported by U.S. DOE grant  DEFG02-90-ER-40560 and NSF grant
PHY900036
}\address{Enrico Fermi Institute,
University of Chicago,
Chicago, IL 60637, U.S.A.}
}

\abstract{
Reductive W-algebras which are generated by bosonic fields of spin-1,
a single spin-2 field and fermionic fields of spin-3/2
are classified. Three new cases are found: a `symplectic'
family of superconformal algebras which are extended by $su(2)\oplus sp(n)$,
an $N=7$ and an $N=8$ superconformal algebra. The exceptional cases can be
viewed as arising a Drinfeld-Sokolov type reduction of the exceptional
Lie superalgebras $F(4)$ and $G(3)$, and have an octonionic description.
The quantum versions of the superconformal algebras
are constructed explicitly in all three cases. }

\maketitle

\section{Introduction}

The construction of superconformal algebras has a long history
\cite{Ade,RSch}.
Because of their easy applicability to string theory,
the first such algebras
were linear and contained one spin-$2$ field $L(z)$,
$N$ spin-$3/2$ fields $G^i(z)$, a number of free fermions $\psi^j(z)$
and spin-$1$ currents $T^a(z)$.
The linearity of the algebra ensured the existence of a finite dimensional
Lie superalgebra consisting of the modes
$L_{-1},L_0,L_1,G^i_{-1/2},G^i_{1/2},T^a_0$.
Since simple superalgebras have been classified, this enabled the
authors of \cite{RSch} to list various possibilities. For a long time
it was thought that this list was complete, until a new $N=4$ algebra was
constructed \cite{STPro,GPTvP} based on a non-simple superalgebra.

Besides these linear algebras,
two families of non-linear superconformal algebras
exist for arbitrary $N$ \cite{Kni,Ber,KSch}. As W-algebras, these are generated
by fields of spin-$1,3/2$ and one field of spin-$2$ but the
anticommutator of two supercharges contains a piece quadratic in the currents
\eq
\{G^i_r,G^j_s\}= ....+\pi^{ab}_{ij} \nox T^a T^b \nox_{r+s}
\label{eq.intro}
\en
where $\nox T^a T^b \nox$ is the normal ordered product of two currents, and
$\pi$ is symmetric in its upper and lower indices.
The vacuum preserving algebra generated by the modes
$L_{-1},L_0,L_1,G_{-1/2},G_{1/2},T^a_0$ is not finite because of the
non-linearity
of the algebra, but can be made so by taking the limit in which the
central charge $c\to \pm\infty$ \cite{BWat}. In this way we can again utilise
Lie superalgebra theory to classify algebras which are associative in this
limit. For instance the two known families of non-linear algebras correspond
to $osp(N,2)$ and $su(N|2)$, or in the notation of Kac \cite{Kac}, $D(N,1)$,
$B(N,1)$ and $A(N,1)$.
{}Furthermore, these algebras arise from a Drinfeld-Sokolov reduction \cite{DS}
of the affine versions of these superalgebras.

It is natural to ask whether or not we can construct any other superconformal
algebras. In particular, the $N=1,2$ and $4$ algebras can be
given a Kazama-Suzuki construction \cite{KSuz} using symmetric,
hermitian symmetric, and quaternionic spaces \cite{vPro}, and this suggests a
natural association with the reals, complex numbers and quaternions,
and in turn suggests the
existence of octonionic analogues for these algebras.
{}Further evidence is provided by the existence of an
octonionic string solution to low energy heterotic string theory \cite{HStr}.
It is tempting to conjecture the existence of some sort of octonionic
superconformal world-sheet symmetry for the effective action of this solution.

In this paper we first classify the possible field contents of
superconformal algebras using the results of \cite{BWat} and \cite{Kac}.
We do indeed find two octonionic superconformal algebras based on the
exceptional superalgebras $F(4)$ and $G(3)$ \cite{SNRit,FKap}, and also new is
a third infinite family extended by $su(2)\oplus sp(N)$.
The main part of this paper is the explicit construction of the quantum
version of each of these algebras by solving Jacobi's identity.
The structure constants of both $F(4)$ and $G(3)$, and thus of the
$N=7$ and $N=8$ algebras that we construct, can be naturally
expressed in terms of octonions \cite{Sud}. It seems likely that the
$N=7$ and $N=8$ algebras are an octonionic
generalisation of the $N=3$ and $N=4$ algebras respectively.
Unfortunately, we find that the symplectically extended
superconformal algebras cannot have unitary highest weight representations.

We conclude with some remarks on future directions of research, and in
particular on the possibility of finding a linear version of these algebras.

\section{Classification of superconformal algebras}

In this section we shall classify $W$-algebras with the following properties.
We assume that the bosonic generators of the algebra consist of
a number of spin-$1$
fields and a spin-$2$ field forming the semi-direct product of a Kac-Moody
algebra with a Virasoro algebra. The fermionic generators are all of
spin-$3/2$.
We exclude the possibility of spin-$1/2$ generators.

We further demand that the
algebra be {\it reductive} \cite{BWat}. This essentially means that the algebra
exists in the limit that the central charge $c\to \infty$ and that the
classical analogue of the algebra can be recovered in this limit.
In this case this implies that there is a coefficient in front of the
quadratic piece of (\ref{eq.intro}) which vanishes in this limit, so that
the algebra linearises.
In \cite{BWat} it was demonstrated that such algebras were characterised
by pairs $(g, h)$ where $g$ is simple (super)algebra and $h$ is an
$su(2)$ subalgebra of $g$. The number, statistics and
conformal dimension of generators
of the corresponding $W$-algebras are given by the number, grading and spins of
irreducible representations in the decomposition of $g$ with respect to $h$.
Each irreducible representation of spin $j$ corresponds to a primary
generator of conformal dimension $j+1$, and bosonic(fermionic) generators
correspond to representations in the even(odd) parts of the superalgebra.

Denote the even and odd parts
of $g$ by $g_E$ and $g_O$. The requirement that the algebra be superconformal,
that is have the field content alluded to above, means that $g_E$ should
decompose into a number of spin-$0$ and one spin-$2$
representation, and $g_O$ into a number of spin-$1/2$ representations
under the decomposition with respect to $h$. In particular, this
implies that $g_E=h\oplus h'$
and that $g_O$ transforms as a $(1/2,\Lambda)$ with respect to $h\oplus h'$,
where $\Lambda$ is some representation of $h'$.

In table 1, we enumerate the possibilities, using the classification of
simple Lie superalgebras \cite{Kac}. In the first column we list the simple
superalgebras. The corresponding even graded algebra $g_E$ is given in the
second column while the next column shows how $g_O$ transforms under $g_E$.
In the last three columns we list the choices of $h$ which yield a
superconformal algebra, the number of supersymmetry generators and the
Kac-Moody component of the resulting algebra. We do not include the
superalgebras $P(n),Q(n),W(n),S(n),\tilde{S}(n)$
or $H(n)$ since these do not yield superconformal algebras, except in
the special cases that they are isomorphic to a superalgebra already
considered.
There is no need to consider semi-simple algebras, since if $h$ is embedded
non-trivially in more than one ideal of $g$ then it is easy to see that the
resulting algebra contains more than one spin-2 field. If, on
the other hand, $h$ is a
subalgebra of some simple ideal in $g$, then the resulting W-algebra is
simply a direct product of the W-algebra associated with the reduction of
this ideal and a Kac-Moody algebra. It is possible to couple these
two parts non-trivially using fields of less than spin-1 such as free
fermions, but this case has been excluded by our assumptions. This possibility
can lead to interesting algebras, as in the case of the linear $N=4$ algebra,
and we return to this point in our conclusions.

\begin{center}
\begin{tabular}{|c|c|c|c|c|c|}
\hline \topspace
$g$&$g_E$& $g_O$ & $(m,n)$ & $N$ & KM \\
\hline \topspace
  $A(m,m)$& $ A_m\oplus A_m$ & $(m+1,m+1)$ & $m=1$
& $4$ & $A_1$
\bottomspace \\
  $A(m,n)$& $u(1)\oplus A_{n+1}$ &$(2m+2,n+1)$&$ n=1$
& $2(m+1)$ & $u(m+1)$
\bottomspace \\
  $B(m,n)$& $B_m\oplus C_n$ &$(2m+1,2n)$&$n=1$
&$2m+1$ & $B_m$
\bottomspace \\
  $D(m,n)$& $D_m\oplus C_n$ &$(2m,n)$& $n=1$
&$2m$ & $D_m$
\bottomspace \\
          &                     &        & $m=2$
&$4m$ & $A_1\oplus C_n$
\bottomspace \\
  $C(n)$& $u(1)\oplus C_n$ &$(-,n)$& $n=1$
&$2$ &$u(1)$
\bottomspace \\
  $D(2,1,\alpha)$& $A_1\oplus A_1 \oplus A_1$ &$(2,2,2)$&
& $4$ &$A_1\oplus A_1  $
\bottomspace \\
  $F(4)$ & $A_1\oplus B_3$ & $(2,8)$
&& $8$ & $B_3$
\bottomspace \\
  $G(3)$ & $A_1\oplus G_2$ & $(2,7)$
&& $7$ & $G_2$
\bottomspace \\
\hline\noalign{\medskip}
\multicolumn{6}{c}{\hbox{Table 1 }}
\end{tabular}
\end{center}
\vskip1cm

One readily identifies $N=1$ and $N=2$ superconformal algebras with
$B(0,1)$, $A(0,1)\equiv C(1)\equiv D(1,1)$, the $N=3$ algebra with $B(1,1)$,
the $N=4$ algebra extended by $su(2)$ with
$A(1,1)$, and
extended by $su(2)\oplus su(2)$ with $D(2,1,\alpha)$.
As mentioned above, the two infinite families
of Knizhnik and Bershadsky can be identified with $A(m,1)$ for $m>1$ and with
$B(m,1)$, $m>1$ and $D(m,1)$, $m>2$. There are three further possibilities:
an infinite family of $su(2)\oplus sp(m)$ extended superconformal algebras
corresponding to $D(2,m)$,
and two exceptional superconformal algebras
corresponding to the exceptional superalgebras $G(3)$ and $F(4)$ \cite{dWvN}.
{}For these algebras one finds that $g_E=su(2)\oplus g_2, su(2)\oplus so(7)$
respectively, and $g_O$ transforms under $g_E$ as a $(1/2,7)$, $(1/2,8)$
respectively. Thus choosing $h$ to be the $su(2)$ factor of $g_E$, we
expect that there exists an $N=7$ superconformal algebra with $\hat{g}_2$
as a subalgebra associated with $G(3)$, and an $N=8$ superconformal algebra
with $\hat{so}(7)$ as a subalgebra associated with $F(4)$.

Though in general it is not known whether the algebra associated to a
particular pair $(g,h)$ is unique, in the next section we shall explicitly
demonstrate that the above algebras are completely determined by their
field content and associativity,  proofing the completeness of our
classification of reductive superalgebras generated by
fields of spin-1, spin-3/2 and spin-2.

\section {The commutation relations of the N=7, N=8  and symplectic
superconformal algebras}

In what follows we derive the commutation relations for the
three new superconformal
algebras based on $G(3)$, $F(4)$ and $D(2,N)$. This simply amounts to
writing down
the general form of the commutation relations and checking Jacobi's identity.
The bosonic part of the algebra consists of the semi-direct product of
a Kac-Moody algebra with the Virasoro algebra:
\eqq
{}~[T^a_m,T^b_n] &=& -f^{abc} T^c_{m+n}+km\delta_{m+n,0} \label{eq.km}
\\ ~[L_m,L_n] &=& (m-n)L_{m+n}+{c\over 12}m(m^2-1)\delta_{m+n,0}
\label{eq.vir}
\\
{}~[L_m,T^a_n]&=&-nT^a_{m+n}
\label{eq.hw}
\enn
where $f^{abc}$ are the structure constants of $g_2$, $so(7)$ and
$su(2)\oplus sp(N)$ for the $N=7$, $N=8$ and symplectic series of algebras
respectively, normalised so that the length of a
long root squared is two. Jacobi's identity is easily verified
for these generators. The superconformal generators $G^i_r$ are highest
weight both for the Kac-Moody algebra and the Virasoro algebra; that is
\eqq
{}~[L_m,G^i_r] &=& [{m\over 2}-r]G^i_{m+r}\\
{}~[T^a_m,G^i_r] &=& M^a_{ij}G^j_{m+r}
\enn
where $M^a_{ij}$ is the seven-dimensional representation of $g_2$, the
eight-dimensional spinor representation of $so(7)$, or the $(2,2N)$
representation of $su(2)\oplus sp(N)$ for the three cases respectively,
and satisfies
\eq
{}~[M^a, M^b]= f^{abc} M^c .
\en
With these commutation relations it is easy to see that the $TTG$ and
$LLG$ Jacobi constraints are satisfied. The only remaining equations that
a consistent algebra must satisfy come from the combinations $TGG$, $LGG$
and $GGG$. To check the first two relations one can exploit Virasoro and
Kac-Moody Ward identities. If we write the operator product expansion
of two superconformal generators as
\eq
G^i(z)G^j(\zeta) = \sum_{n\geq 0} \psi^{ij}_n(\zeta) (z-\zeta)^{-3+n},
\label{eq.superope}
\en
then these identities are respectively
\eqq
L_m|\psi^{ij}_n\rangle &=& (n+m/2-3/2)|\psi^{ij}_{n-m}\rangle
\label{eq.vw}\\
T^a_m|\psi^{ij}_n\rangle &=& M^a_{ik}|\psi^{kj}_{n-m}\rangle,
\label{eq.kmw}
\enn
where $n>0$ and the state $|\psi\rangle$ is given by the usual correspondence
$|\psi\rangle=\lim_{z\to 0}\psi(z)|vac\rangle$. Once these two relations
have been satisfied, it only remains to check the $GGG$ Jacobi identity.
We now deal with each case in turn. The calculation is somewhat involved,
and a number of relevant identities and conventions are given in an
appendix.

\subsection{The $N=7$ superalgebra}

Writing out explicitly the operator product expansion (\ref{eq.superope})
in modes and in a basis of primary fields we find that the most general
form for the anticommutation relation of two superconformal generators
can be written
\eqq
\{G^i_r,G^j_s\} &=& {c\over 3}(r^2-{1\over 4})\delta_{r+s,0}
+B\delta^{ij}L_{r+s} +\delta^{ij}D(L_{r+s}-{c\over {c_g}}{\cal L}_{r+s})
\nonumber\\
              &+& C{{r-s}\over 2} M^a_{ij}T^a_{r+s}+E\pi^{ab}_{ij}
(\nox T^a T^b \nox)_{r+s}
\label{eq.supercom}
\enn
where $B,C,D,E$ are coefficients to be determined, ${\cal L}$ is the
Sugawara energy momentum tensor
\eq
{\cal L}_m={1\over 2(k+4)}(\nox T^a T^a \nox)_m
\en
for $g_2$ whose central charge is given by $c_g= 14k/(k+4)$.
The normalisation of $G$ has been fixed by the first term in
(\ref{eq.supercom}). The tensor $\pi^{ab}_{ij}$ is symmetric and traceless
in both upper and lower indices. The Virasoro Ward identity (\ref{eq.vw})
gives that $B=2$. The Kac-Moody Ward identity implies that $C=2c/3k$ and
yields the equation
\eqq
{2c\over 3k} M^a_{ix} M^b_{xj} &=&
{c\over 3k} f^{abc}M^c_{ij}+ \delta^{ab}\delta_{ij}[2+D(1-{c\over {c_g}})]
\nonumber\\
&+& \pi^{ef}_{ij}[2k\delta^{ea}\delta^{fb}-f^{yae}f^{ybf}]
\label{eq.ntkm}
\enn
The matrix $M^a$ satisfies the properties
\eqq
Tr (M^a M^b) &=&2\delta^{ab}\\
\sum_a M^a_{ij} M^a_{kl} &=& {2\over 3}(\delta_{jk}\delta_{il}-
\delta_{ik}\delta_{jl})+1/3\eta_{ijkl}
\label{eq.outerproduct}
\enn
where the $\eta$ is a totally antisymmetric matrix best defined in terms
of the octonion structure constants $c_{ijk}$ as
\eq
c_{xij}c_{xkl}=\eta_{ijkl}+\delta_{ik}\delta_{jl}-\delta_{il}\delta_{jk}.
\en
{}For a particular choice for the values of $c_{ijk},\eta_{ijkl}$, and $M$
we refer the reader to the appendix and to \cite{SNRit}.
Taking the trace of (\ref{eq.ntkm}) we find immediately that
\eq
2+D(1-{c\over {c_g}}) ={4c\over 21k} .
\label{eq.D}
\en
The traceless symmetric part of (\ref{eq.ntkm}) is solved by setting
\eqq
\pi^{ab}_{ij} &=& (M^a M^b + M^b M^a)_{ij}-{4\over 7}\delta^{ab}\delta_{ij}\\
E{{6k+10}\over 3}&=&{c\over 3k}
\label{eq.E}
\enn
where we have made extensive use of (\ref{eq.outerproduct}).

Our only remaining task is to ensure that the Jacobi identity for three
superconformal generators is satisfied. Noting that
\eqq
{}~[\nox T^a T^b \nox_m, G^i_r]
&=& {1\over 3} ({m\over 2}-r)(M^a M^b + M^b M^a )_{ij}G^j_{m+r}
\nonumber\\
&-& (m+1)f^{abc}M^c_{ij}G^j_{m+r}
+M^a_{ij}\nox T^b G^j \nox_{m+r}+M^b_{ij}\nox T^a G^j \nox_{m+r}
\enn
and using a number of other identities (again in the appendix) we find
two further equations
\eqq
{32E\over 21} &=& {cD\over {c_g(k+4)}}
\label{eq.JI1}\\
2+D\left (1-{4c\over {3c_g(k+4)}}\right ) &=& {4c\over 9k}+{32E\over 21}
\label{eq.JI2}
\enn

Equations (\ref{eq.D}),(\ref{eq.E}),(\ref{eq.JI1}) and (\ref{eq.JI2})
contain the four variables $D,E,k$ and $c$. This is in seeming contradiction
to the assumption
that this algebra
is reductive; that is to say it is associative for continuous $c$.
Thus we expect that the value of
$c$ is not fixed and parameterises the solution.
It is a good check on our calculation that indeed the
four equations are not independent, and admit the following one parameter
family of solutions
\eqq
c &=& {k(9k+31)\over 2(k+3)}\\
E &=& {c\over 2k(3k+5)}\\
D &=& {32\over 3(3k+5)}
\enn

As mentioned earlier, we expect that $W$-algebras of this type
linearise in the limit that $c\to \infty$. In this limit $c$ is proportional
to $k$
and we note that
the coefficients $D,E$ of the quadratic terms in (\ref{eq.supercom}) tend to
zero, confirming our expectations.

The superalgebra $G(3)$ has an octonionic description in which the bosonic
part is viewed as the direct sum of $su(2)$ and the derivation algebra
$g_2$ of the octonions. The fermionic part can then be represented as a
doublet of pure imaginary octonions \cite{Sud}. The quaternionic version
of this algebra has the direct product of
an $su(2)$ with the derivation algebra of the quaternions which is another
copy of $su(2)$ as its bosonic part,
acting on a doublet of the three pure imaginary quaternions.
This is the superalgebra $B(1,1)$ in Kac's notation which is the superalgebra
associated with the $N=3$ superconformal algebra. Similarly we can associate
the Virasoro algebra and the $N=1$ superconformal algebra with the reals and
the complex numbers.

\subsection{The $N=8$ superconformal algebra}

The calculation for the $N=8$ superconformal algebra is very similar to
those sketched above. It is more convenient to label the adjoint representation
of $so(7)$ by a pair of antisymmetric indices $I,J=1...7$, so that the
commutation relations of the corresponding Kac-Moody algebra are
\eqq
{}~[T^{IJ}_m,T^{KL}_n] &=& -i(
\delta_{JK}\delta_{IM}\delta_{LN}
-\delta_{IK}\delta_{JM}\delta_{LN}
-\delta_{JL}\delta_{IM}\delta_{KN}
+\delta_{IL}\delta_{JM}\delta_{KN})T^{MN}_{m+n}\nonumber\\
&+& km(\delta^{IK}\delta^{JL}-\delta^{IL}\delta^{JK})
\enn
Since we are interested in the spinor representation of $so(7)$ we
introduce gamma matrices which can be written \cite{CDFN}
\eq
(\gamma^I)_{ij}=i(c_{Iij}+\delta_{Ii}\delta_{j8}-\delta_{Ij}\delta_{i8})
\label{eq.gamma}
\en
where $c_{ijk}$ are the structure constants of the octonions as above.
We also adopt the normal convention
$\gamma^{IJ ..N}=\gamma^{[I}\gamma^J..\gamma^{N]}$.
The most general form for the commutator of two superconformal generators
is given by
\eqq
\{G^i_r,G^j_s\} &=& {c\over 3}(r^2-{1\over 4})\delta_{r+s,0}
+B\delta^{ij}L_{r+s} +\delta^{ij}D(L_{r+s}-{c\over {c_g}}{\cal L}_{r+s})
\nonumber\\
              &+& C{{r-s}\over 2} \gamma^{IJ}_{ij}T^{IJ}_{r+s}
+E\gamma^{IJKL}_{ij}(\nox T^{IJ} T^{KL} \nox)_{r+s}
\label{eq.supercom8}
\enn
The Virasoro and Kac-Moody Ward identities give that
\eqq
B &=& 2\\
C &=& {ic\over 6k}\\
2+D(1-{c\over {c_g}})&=&{c\over 6k}
\label{eq.D8}\\
E&=&-{c\over 48k(k+2)}
\label{eq.E8}
\enn
The Jacobi identity for three superconformal generators gives us the two
equations
\eqq
{cD\over {c_g(4k+20)}}&=&-4E
\label{eq.JI18}\\
2+D(1-{c\over {c_g(4k+20)}})-{c\over 2k}+20E &=& 0.
\label{eq.JI28}
\enn
Again we find that the four equations (\ref{eq.D8}),(\ref{eq.E8}),
(\ref{eq.JI18}) and (\ref{eq.JI28}) are not independent and have the
solution
\eqq
c &=& {2k(2k+11)\over (k+4)}\\
D &=& {7\over (k+2)}\\
E&=&-{c\over 48k(k+2)}.\
\enn

As in the $N=7$ case, this algebra has a natural description in
terms of the structure constants of the octonions. It is not
entirely obvious what the quaternionic analogue of this algebra is.
The most naive guess might be an algebra with four supercharges
transforming under the spinor and complex conjugate spinor representations
of $so(3)$, or in other words, the superconformal algebra of \cite{Ade}.
On the other hand one can introduce an extra $so(3)$ into the quaternionic
analogy by considering the group of similarity triples \cite{Sud,Ram}, rather
than the group of norm preserving transformations, suggesting that we should
consider the non-linear $\tilde{A}_\gamma$ algebra as the quaternionic
equivalent.

\subsection{The symplectic superconformal algebras}

Corresponding to the superalgebras $D(2,N)$ there exists a family of
superconformal algebras with $4N$ fermionic generators. The algebra
contains an $su(2)\oplus sp(N)$ Kac-Moody component, under which
the fermionic generators transform as a $(2,2N)$ representation. We
label $sp(n)$ adjoint indices by $x,y...$ and $su(2)$ adjoint
indices by $a,b...$. We label the $4N$-dimensional representation
space by Greek letters $\alpha, \beta...$ although it will be
convenient to decompose these indices into a pair of indices $(i,I)$
where $i=1,...4$ and $I=1,...N$. Thus we write
\eqq
{}~[T^a_m,G^\alpha_r]&=&M^a_{\alpha \beta}G^\beta_{m+r}\\
{}~[T^x_m,G^\alpha_r]&=&M^x_{\alpha \beta}G^\beta_{m+r}
\enn
where we can take the generators of $su(2)$ to be
$ {i\over \sqrt{2}}\gamma^{+a}_{ij}\delta_{IJ}$ and the generators
of $sp(N)$ can be taken to be
${i\over \sqrt{2}}\gamma^{-a}_{\alpha\beta}{\delta_{AI}\delta_{AJ}}$
for $A=1..N$,
$i\gamma^{-a}_{\alpha\beta}(\delta_{AI}\delta_{BJ}-\delta_{AJ}\delta_{BJ})$
and
${i\over
2}\delta_{\alpha\beta}(\delta_{AI}\delta_{BJ}-\delta_{AJ}\delta_{BJ})$.Here we
can take
\eq
\gamma^{a\pm}_{\alpha\beta}=\epsilon_{a\alpha\beta}\pm(\delta_{a\alpha}
\delta_{\beta 4}-\delta_{a\beta}\delta_{\alpha 4}).
\en
Note that this is simply the quaternionic analogue of the formula
(\ref{eq.gamma}). With this choice $tr(M^a M^b)=2N\delta^{ab}$, $tr(M^a M^x)=0$
and $tr(M^x M^y)=2\delta^{xy}$. Unlike the previous examples, the Kac-Moody
part of the algebra is the direct sum of two simple pieces, and so the
most general form of the commutation relations contain the central charge
and {\it two} levels $k,k'$ where
\eqq
{}~[T^a_m, T^b_n]&=& -f^{abc}T^c_{m+n}+km\delta^{ab}\delta_{m+n,0}\\
{}~[T^x_m, T^y_n]&=& -f^{xyz}T^z_{m+n}+k'm\delta^{xy}\delta_{m+n,0}
\enn
and we have normalised the currents so that $[M^a, M^b]=f^{abc}M^c$ etc.
as before.

The anticommutator of two fermionic generators can be written as
\eqq
\{G^{\alpha}_r,G^{\beta}_s\} &=& {c\over 3}(r^2-{1\over 4})\delta_{r+s,0}
+2\delta^{{\alpha}{\beta}}L_{r+s}
\nonumber\\
&+&\delta^{{\alpha}{\beta}}
D(L_{r+s}-{c\over {c_g}}{\cal L}^{su(2)}B_{r+s})+
D'(L_{r+s}-{c\over {c_g'}}{\cal L}^{sp(n)}_{r+s})
\nonumber\\
&+& {{r-s}\over 2} ({2c\over 3k}M^a_{{\alpha}{\beta}}T^a_{r+s}+
{2c\over 3k'}M^x_{\alpha\beta}T^x_{r+s})
\nonumber\\
&+&E\pi^{ax}_{{\alpha}{\beta}}
(\nox T^a T^x \nox)_{r+s}
+E'\pi^{xy}_{{\alpha}{\beta}}
(\nox T^x T^y \nox)_{r+s}
\label{eq.supercomsym}
\enn
where
\eqq
\pi^{xy}_{\alpha\beta} &=& (M^aM^b+M^bM^a)_{ \alpha\beta}-1/N\delta^{xy}
\delta_{\alpha\beta}\\
\pi^{ab}_{\alpha\beta} &=& (M^xM^y+M^yM^x)_{ \alpha\beta}\\
c_g&=&{3k\over k+2}\\
c_{g'}&=&{N(2N+1)k'\over {k'+N+1}}.
\enn
Note that there is no term containing the symmetric traceless product
of currents $\nox T^a T^b \nox$. This is because such a product transforms as
spin-$2$ multiplet under $su(2)$ and there is no invariant way to couple
this to the pair of spin-$1/2$ indices $\alpha,\beta$.

Demanding associativity yields the following equations:
\eqq
E &=& {c\over 3kk'} \label{eq.Esym}\\
E'&=& {c\over {3k'(2k'+N+2)\label{eq.E'sym}}}  \\
2+D(1-{c\over c_g})+D' &=& {c\over 3k} \\
2+D'(1-{c\over c_{g'}})+D &=&  {c\over 3k'N}\\
E'{N+1\over 4}-{E\over 4} &=& {c\over 6k}+{c\over 12k'}\label{eq.Dsym}\\
{1\over 2}[2+D+D'-{cD\over 2c_g(k+2)}&\ &\\
-{cD'(2N+1)\over 6c_{g'}(k'+N+1)}]
&=& {c\over 12k'}+E'{(N+1)(N+2)\over 12N}+{E\over 4}\label{eq.D'sym}\\
E &=& -E'\label{eq.E''sym}\\
{cD'\over c_{g'}(k'+N+1)} &=& E'{(N-2)\over N}\label{eq.Fsym}\\
{cD\over c_{g}(k+2)} &=& E.
\enn
{}From these it is easy to show that the levels $k,k'$ are related by the
equation
\eq
2k'+k+N+2=0
\label{eq.levels}
\en
The remaining variables are given by
\eqq
D&=&{1\over k'}\\
D' &=& {-(N-2)(2N+1)\over 3k}\\
E &=& -E' = {c\over 3kk'}\\
c &=& {{6kk'+3k-k'(N-2)(2N+1)}\over {k+k'+2}}.
\label{eq.cent}
\enn

As a check we note that for $N=1$ this algebra should coincide with
the known non-linear $N=4$ algebra, $\tilde{A}_{\gamma}$. The formula
for the central charge of $\tilde{A}_\gamma$ is
\eq
c= {6(k+1)(k'+1)\over k+k'+2}-3
\en
which coincides with (\ref{eq.cent}) for $N=1$.
The formulae for the coefficients $D,D'$ are only correct if we use the
relation between the levels (\ref{eq.levels}) which is not needed for
associativity of the $\tilde{A}_\gamma$ algebra. In fact one needs to rederive
the associativity equations in this case since a number of special relations
hold. For instance, the tensor $\pi^{xy}$ identically vanishes.
coefficient is $E'$ vanishes
Unfortunately, for $N>1$
there is no possibility that this algebra can have unitary
highest weight representations since (\ref{eq.levels}) does not allow $k$
and $k'$ to be positive at the same time, which is a requirement for
such representations.

\section{Conclusions}

We have succeeded in constructing three new non-linear quantum
superconformal algebras,
based on the Lie superalgebras $F(4)$, $G(3)$ and $D(2,m)$. One can easily
recover the
classical versions of these algebras by making the substitutions
\eq
L'=\hbar L\;,\; G'^i=\hbar G^i\;,\;T'^a=\hbar T^a, c'=\hbar c
\en
and calculating the Poisson brackets of primed quantities using the
usual correspondence; e.g.
\eq
\{G'^i,G'^j\}_{{\rm P.B}}=\lim_{\hbar\to 0}\{G'^i,G'^j\},
\en
while keeping $c'$ fixed.

The symplectic series of superconformal algebras do not seem useful, at least
in the context of rational conformal
field theory, since they do not admit highest weight unitary representations.
The $N=7$ and $N=8$ superalgebras have no obvious obstruction to having
such representations however and it should be interesting to develop
a representation theory for these algebras. We also need to make the analogy
between these algebras and the
non-linear $N=3$ and $N=4$ algebras more precise.
The latter can be constructed using a Kazama-Suzuki construction
on quaternionic(Wolf) spaces \cite{vPro}. It would be interesting to see
whether
some corresponding construction exists for the new $N=8$ algebra, and
what space one would need to consider in such a construction.
The free field realisation of these
algebras which arises from a generalised Drinfeld-Sokolov
reduction of the superalgebras $F(4)$ and $G(3)$ may be related to the
above questions.

The classification we give is only complete if we restrict ourselves to
extending the Virasoro algebra by fields of spin-$1$ and spin $3/2$.
{}For instance, both the $N=3$ and the $N=4$ algebras exist in linear
and non-linear forms, so that we may hope that the same is true of their
octonionic versions. It is known how to systematically remove the free
fermions from a linear algebra to obtain a fermion-free and generically
non-linear algebra \cite{GSch}.
However, it is not clear how to implement the reverse
procedure, coupling the non-linear algebras to a number of free
fermions which may interact further with a new Kac-Moody algebra to
obtain a linear algebra.
It remains an open question whether this can be done in the present case.
The analysis of \cite{GSTvP} does not include
the possibility of spin-$1/2$ fermions, and though the authors of
\cite{RSch} include fermions, the situation where these are coupled
to another Kac-Moody algebra is not considered.
We hope to settle this question in the near future.
\blank{
CITATIONS TO GO IN
\cite{BGer2,BGer3,BGer4,GERV,FRRT1,ORTW1,OWip1,BFFOW2,BTDr1,Fehe1,Bers1,BOog1,PRIN,AS}
}

\section{Appendix}

The structure constants $c_{ijk}$ can be taken to be
totally antisymmetric. We shall use the notation of \cite{SNRit} and so
$c_{ijk}$ takes the value one
for any of the triples
\eq
(1,2,3),\;(1,4,5)\;,(1,7,6)\;,(2,4,6)\;,(2,5,7)\;,(3,4,7)\;,(3,6,5).
\nonumber
\en
For this choice the tensor $\eta_{ijkl}$ is non-vanishing for
permutations of
\eqq  (1,2,4,7),\;(1,2,6,5),\;(1,3,6,4),\;(1,3,5,7),\;\nonumber\\
      (2,3,4,5),\;(2,3,7,6),\;(4,5,7,6).
\nonumber
\enn

The relation (\ref{eq.outerproduct}) can be proved by considering
the following embedding of $g_2\in so(7)$:
\eqq
M^1 &=& {1\over \sqrt{2}}(T^{41}+T^{36})\;\;\;\;
M^2  =  {1\over \sqrt{6}}(T^{41}-T^{36}-2T^{27})\nonumber\\
M^3 &=& {1\over \sqrt{2}}(T^{31}-T^{46})\;\;\;\;
M^4  =  {1\over \sqrt{6}}(T^{31}+T^{46}-2T^{57})\nonumber\\
M^5 &=& {1\over \sqrt{2}}(T^{21}-T^{56})\;\;\;\;
M^6  =  {1\over \sqrt{6}}(T^{21}+T^{56}+2T^{47})\nonumber\\
M^7 &=& {1\over \sqrt{2}}(T^{51}+T^{26})\;\;\;\;
M^8  =  {1\over \sqrt{6}}(T^{51}-T^{26}+2T^{37})\nonumber\\
M^9 &=& {1\over \sqrt{2}}(T^{24}-T^{53})\;\;\;\;
M^10  =  {1\over \sqrt{6}}(T^{24}+T^{53}-2T^{17})\nonumber\\
M^{11} &=& {1\over \sqrt{2}}(T^{54}+T^{23})\;\;\;\;
M^{12}  =  {1\over \sqrt{6}}(T^{54}-T^{23}-2T^{67})\nonumber\\
M^{13} &=& {1\over \sqrt{2}}(T^{43}-T^{16})\;\;\;\;
M^{14}  =  {1\over \sqrt{6}}(T^{43}+T^{16}+2T^{25})\nonumber
\enn
where the matrix $T^{IJ}$ can be taken to be the vector representation
\eq
T^{IJ}_{ij}=-i(\delta_{Ii}\delta_{Jj}-\delta_{Ij}\delta_{Ji})
\nonumber
\en
The matrix $\pi^{ab}_{ij}$ satisfies the equations
\eqq
\sum_{_{\rm cyclic}ijk}\pi^{ab}_{ij}M^a_{kl}&=&
{16\over 21}\sum_{_{\rm cyclic}ijk}
\delta_{ij}M^b_{kl}\nonumber\\
\pi^{ab}_{ij}\pi^{ab}_{kl}&=&{32\over 9}
(\delta_{ik}\delta_{jl}+\delta_{jk}\delta_{il}-{2\over7}\delta_{ij}\delta_{kl})
\nonumber
\enn

{}For the $N=8$ case it is useful to extend the tensor $\eta_{ijkl}$ to the
totally antisymmetric tensor \cite{CDFN,WNic}
\eq
c^{\pm}_{abcd}=\eta_{abcd}\mp (c_{abc}\delta_{d8}+c_{bcd}\delta_{a8}
+c_{cda}\delta_{b8}+c_{dab}\delta_{c8})
\nonumber
\en
where now the indices $a,b,c,d$ run from $1$ to $8$.
One can establish that
\eqq
\gamma^{ij}_{ab}&=&c^+_{ijab}+\delta_{ia}\delta_{jb}-\delta_{ib}\delta_{ja}
\nonumber\\
\sum_{ij=1}^7  \gamma^{ij}_{ab}\gamma^{ij}_{cd}&=&
6(\delta_{ac}\delta_{bd}-\delta_{ad}\delta_{bc})-2c^-_{abcd}
\nonumber\\
\sum_{ij=1}^7  \gamma^{ijkl}_{ab}\gamma^{ij}_{cd}&=&
3\delta_{ac}\gamma^{kl}_{db}
-3\delta_{ad}\gamma^{kl}_{cb}
+3\delta_{bd}\gamma^{kl}_{ac}
-3\delta_{bc}\gamma^{kl}_{ad}
\nonumber\\
&+& 2\delta_{ab}\gamma^{kl}_{cd}+c^-_{axcd}\gamma^{kl}_{bx}
+c^-_{bxcd}\gamma^{kl}_{ax}
\nonumber\\
\gamma^{ijkl}_{ab}\gamma^{ijkl}_{cd} &=& 96(\delta_{ac}\delta_{bd}-
\delta_{bc}\delta_{ad})-24\delta_{ab}\delta_{cd}.
\enn

In the symplectic case, most relations needed can be proved using
\eqq
\sum_{x\in sp(N)}M^a_{iIjJ}M^b_{kKlL} =
&-&{1\over 4}\delta_{ij}\delta_{kl}[\delta_{IK}\delta_{JL}
-\delta_{IL}\delta_{JK}] \nonumber\\
&-&{1\over 4}[\delta_{ik}\delta_{jl}-\delta_{jk}\delta_{il}-\epsilon_{ijkl}]
[\delta_{IK}\delta_{JL}+\delta_{IL}\delta_{JK}]
\nonumber\\
\sum_{a\in su(2)}M^x_{iIjJ}M^x_{kKlL} =
&-&{1\over 2}[\delta_{ik}\delta_{jl}-\delta_{jk}\delta_{il}+\epsilon_{ijkl}]
\delta_{IJ}\delta_{KL}\nonumber
\enn
The matrices $\pi^{xy}$ and $\pi^{ax}$ satisfy the equations
\eqq
\sum_{_{\rm cyclic}\alpha\beta\gamma}2\pi^{xy}_{\alpha\beta}
M^y_{\gamma\delta}-\pi^{xa}_{\alpha\beta}M^a_{\gamma\delta}&=&
{N-2\over N}\sum_{_{\rm cyclic}\alpha\beta\gamma}
\delta_{\alpha\beta}M^x_{\gamma\delta}\nonumber\\
\sum_{_{\rm cyclic}\alpha\beta\gamma}2\pi^{ay}_{\alpha\beta}
M^y_{\gamma\delta}&=&
\sum_{_{\rm cyclic}\alpha\beta\gamma}
\delta_{\alpha\beta}M^a_{\gamma\delta}\nonumber
\enn

\end{document}